\documentclass[aps, prl, a4paper, showpacs, twocolumn, english, 10pt]{revtex4-1}
\usepackage{bbm, amsthm, bm,textcomp, nicefrac,geometry,ragged2e}
\usepackage[dvipsnames]{xcolor}
\usepackage{float}
\usepackage[bbgreekl]{mathbbol}
\usepackage{graphicx,epstopdf,color,verbatim,enumitem,ulem}
\usepackage{pbox,array}
\usepackage{mathrsfs}
\usepackage[dvipsnames]{xcolor}
\usepackage[export]{adjustbox}
\usepackage{amssymb}
\usepackage{mathtools}
\usepackage{stackrel}
\usepackage[thinlines]{easytable}
\usepackage{amsmath}
\makeatletter
\usepackage{hyperref}
\usepackage[euler]{textgreek}
\hypersetup{
    colorlinks = true,
    linkcolor = Black,
    citecolor = Black,
    filecolor = Black,      
    urlcolor = Black,
}
\usepackage[caption=false]{subfig}
\usepackage[T1]{fontenc}
\usepackage[caption=false]{subfig}
\usepackage{babel}
\addto\captionsenglish{}

\newcommand\id{\mathbb{1}}

\newcommand{\be}{\begin{equation}}
\newcommand{\ee}{\end{equation}}
\newcommand{\eea}{\end{eqnarray}}
\newcommand{\bea}{\begin{eqnarray}}

\definecolor{brickred}{rgb}{0.8, 0.0, 0.0}

\begin{document}
\title{Entanglement-Assisted Entanglement Purification}
\author{F. Riera-S\`abat$^1$, P.~Sekatski$^2$, A.~Pirker$^{1}$ and W.~D\"ur$^1$}
\affiliation{$^1$Institut f\"ur Theoretische Physik, Universit\"at Innsbruck, Technikerstra{\ss}e 21a, 6020 Innsbruck, Austria\\
$^2$Departement Physik, Universit\"at Basel, Klingelbergstra{\ss}e 82, 4056 Basel, Switzerland}
\date{\today}
\begin{abstract}
The efficient generation of high-fidelity entangled states is the key element for long-distance quantum communication, quantum computation, and other quantum technologies, and at the same time the most resource-consuming part in many schemes. We present a class of entanglement-assisted entanglement purification protocols that can generate high-fidelity entanglement from noisy, finite-size ensembles with improved yield and fidelity as compared to previous approaches. The scheme utilizes high-dimensional auxiliary entanglement to perform entangling nonlocal measurements and determine the number and positions of errors in an ensemble in a controlled and efficient way, without disturbing the entanglement of good pairs. Our protocols can deal with arbitrary errors, but are best suited for few errors, and work particularly well for decay noise. Our methods are applicable to moderately sized ensembles, as will be important for near term quantum devices.
\end{abstract}
\maketitle

{\it Introduction.---}
Entanglement plays a dominant role in most applications of quantum technologies, ranging from metrology \cite{RevModPhys.90.035005,giovannetti2011advances,T_th_2014} and quantum computation \cite{dur2003entanglement,campbell2008measurement, li2012high,zwerger2014robustness,raussendorf2001one,briegel2009measurement} to quantum communication. Quantum advantages in metrology are based on using entangled states, and there exist models for quantum computation where information processing takes place solely by performing measurements on universal resource states \cite{raussendorf2001one,zwerger2014robustness,briegel2009measurement}, or certain gates are implemented in a fault-tolerant fashion by consuming (magic) entangled states \cite{campbell2017roads,PhysRevA.71.022316,knill2005quantum}. Most importantly, quantum networks \cite{kimble2008quantum,wehner2018quantum} and long-distance quantum communication are based on the generation of high-fidelity entanglement using quantum repeaters \cite{briegel1998quantum,dur1999quantum,Ladd_2006,RevModPhys.83.33,sheng2013hybrid,azuma2015all,zwerger2016measurement,zwerger2018long,pirandola2017fundamental,pirandola2019end}. The key elements in all these schemes are methods to generate and maintain high-fidelity entanglement. Entanglement purification protocols (EPPs) \cite{bennett1996purification, bennett1996mixed,deutsch1996quantum,dur2007entanglement,yamamoto2001concentration,SDP-bound,dehaene2003local,vollbrecht2005interpolation,bombin2005entanglement,bombin2006topological,fujii2009entanglement,buscemi2010distilling,brandao2011one,zwerger2013universal,bratzik2013quantum,nickerson2014freely,ruan2018adaptive,rozpkedek2018optimizing,wallnofer2019multipartite,krastanov2019optimized,fang2019non,de2020protocols,fang2020no,fang2020no2,hu2021long,Zhou:20} achieve this aim by locally manipulating ensembles of multiple copies of noisy entangled states in such a way that fewer copies with increased fidelity are produced. The yield and the reachable fidelity of these schemes determine the performance of higher-level applications, as EPPs are used as elementary blocks that are e.g., in the case of quantum repeaters, iteratively applied and, in fact, give rise to the major part of the required resources \cite{briegel1998quantum,dur2007entanglement,dur1999quantum}. Improving the efficiency of these schemes has hence a huge impact on the performance of entanglement-based quantum technologies.

Here we introduce a class of entanglement-assisted entanglement purification protocols with improved yield and reachable fidelity. The basic idea is to use auxiliary, high-dimensional entangled states to read out locally inaccessible information about an ensemble of noisy entangled states, thereby purifying it. The advantage that qudits instead of qubits provide in quantum information theory has already been shown, and many experimental demonstrations regarding the control and generation of entangled \textit{d}-level systems have been reported \cite{cozzolino2019high,erhard2020advances,wang2020qudits,ecker2019overcoming}. In contrast to asymptotic schemes such as hashing or breeding \cite{bennett1996mixed}, the usage of high-dimensional entanglement allows for direct and controlled access to the number, type, and position of errors and the design of efficient schemes that work for an arbitrary finite number of copies. Our error identification protocols (EIPs) interpolate between recurrence protocols \cite{bennett1996purification,bennett1996mixed,deutsch1996quantum,dur2007entanglement} that operate on few copies and are iteratively applied and asymptotic schemes \cite{bennett1996mixed,dur2007entanglement} that operate on large ensembles where partial information is repeatedly accessed and interpreted using laws of large numbers, but outperform both in the important regime of small and moderate number of copies. The EIPs can deal with any noise, including nonindependent identically distributed (IID) situations, but are particularly simple and efficient if the dominant noise process is such as decay. This is the case for many setups, where, e.g., the decay of an excited state of an atom or the loss of a photon (in photon-number entanglement) constitute most relevant processes. Even though the EIPs require auxiliary entanglement, we show how to use only noisy states from the initial ensemble and to deal with errors in operations (in a similar fashion as the breeding protocol leads to hashing). The key ingredient of our protocols is a class of entangling nonlocal measurements, implemented via local operation and classical communication with the help of auxiliary entanglement. This method to read out locally inaccessible information is, however, not limited to EPP, but can in fact be used whenever such ``nonlocal information'' about ensembles should be learned, e.g., for entanglement certification or decision problems.

{\it Setting.---}
We consider two parties sharing an ensemble $\Gamma = \rho^{ \otimes n}$ of $n$ copies of a noisy Bell state $\rho$ with initial local fidelity $F = \langle \Psi_{00} | \rho | \Psi_{00} \rangle$ and global fidelity $F_g =\langle \Psi_{00}|^{\otimes n} \Gamma|\Psi_{00}\rangle^{\otimes n}$. Here we denote by $|\Psi_{ij} \rangle_{AB} = \id \otimes \sigma^j_x \sigma^i_z ( | 00 \rangle_{AB} + |11\rangle_{AB} ) / \sqrt{2}$ the four Bell states, and we consider shared auxiliary maximally entangled states of qudits $|\Psi^{(d)}_{00}\rangle$, where $|\Psi^{ (d) }_{mn} \rangle_{AB} = \sum_{ k = 0 }^{ d - 1 } e^{ i \frac{2\pi}{d} km } |k\rangle_A |k \ominus n \rangle_B / \sqrt{d}$. By performing local operations, the parties aim to purify the ensemble, i.e., probabilistically generate a new ensemble $\Gamma'$ of $2m$ qubits, with increased global fidelity $F'_g$ or local fidelity $F'^{ (k) } = \langle\Psi_{00} | \text{tr}_{\neg k} \Gamma' | \Psi_{00} \rangle$. Usually, we are concerned with reaching a certain global or local target fidelity and define the yield of the procedure as $Y = m'/n$, where $m'$ is the number of final copies that fulfill this fidelity criterion. The prime on $m'$ indicates that auxiliary entanglement that was consumed during the process needs to be restored from the $m$ output pairs. For a better comparison, we assume that $\nu$ pairs with fidelity $F$ correspond to a maximally entangled system with $d = 2^{[ 1 - S (F) ] \nu}$, which corresponds to the (reachable) yield of the asymptotic hashing protocol. Here, $S (F)$ denotes the entropy with respect to diagonal coefficients of the state in the Bell basis, and we use that $|\Psi_{00} \rangle^{ \otimes n } \simeq | \Psi^{ (2^n) }_{00} \rangle$ \cite{horodecki2009quantum}. The EIPs we introduce have the possibility to abort in unfavorable cases, providing higher fidelities or yield.

We consider noisy initial states of the form
\be
    \label{StandardForm}
    \rho = \sum_{ j = 0 }^3 p_j | \phi_j \rangle \langle \phi_j |,
\ee
with $|\phi_j\rangle \in \{|\Psi_{00}\rangle, |01\rangle, |10\rangle, |\Psi_{10}\rangle \}$. Any state can be depolarized, i.e., transformed by applying randomly local operations, to this standard form keeping the fidelity $F = p_0$. The depolarizing procedure (DEP) is given by channels $\mathcal{D}_2 \circ\mathcal{D}_1$ the Kraus operators of which are $\mathcal{D}_1 : \{ \frac{1}{2} \sigma_i \otimes \sigma_i \}_{ i = 0 }^3$ \cite{bennett1996mixed,bennett1996purification,deutsch1996quantum} and $\mathcal{D}_2 : \{ \frac{1}{\sqrt{2}} \id \otimes \id, \frac{1}{ \sqrt{2} } e^{i\frac{\pi}{2} | 1 \rangle \langle 1 | } \otimes e^{ i \frac{\pi}{2} | 0 \rangle \langle 0 | } \}$. We call $| 01 \rangle$, $| 10 \rangle$ and $| \Psi_{ 10 } \rangle$ the error states. An ensemble $\rho^{ \otimes n }$ can be understood as a mixture of (all permutations of) pure states of the form $\bigotimes_{j = 0}^3 | \phi_j \rangle^{ \otimes n_j }$ with $\sum n_j = n$, where the corresponding probabilities follow a multinomial distribution. Our task is to figure out which of these configurations is present, i.e., to identify the position of all error states. This knowledge corresponds to a purification that provides at most $n_0$ Bell states. In contrast to hashing \cite{bennett1996mixed,dur2007entanglement} we do not require to learn the type of error, as the error states that we detect are separable and are discarded.

First we consider rank-2 states with $p_2 = p_3 = 0$,
\be
    \label{eq:toy:state}
    \rho_1 = F |\Psi_{00}\rangle \langle \Psi_{00}| + \big(1 - F \big) |01\rangle \langle 01|.
\ee
Such states are local unitary equivalent to an entangled state $|\Psi_{10}\rangle$ where both qubits are subjected to a decay channel \cite{nielsen2002quantum}, e.g., resulting from the decay of an electronic excited state of atoms or loss of a photon when using a photon-number encoding. This is a relevant class of states, as such noise processes are dominant in many setups.

We introduce a so-called counter gate, which is used to determine the number and position of errors. As this information is locally inaccessible (at least without destroying the entanglement), auxiliary entanglement is required. The counter gate increases [reduces] (mod \textit{d}) the index of an auxiliary entangled state for $|01\rangle$ [$|10\rangle$] error states, while leaving it invariant in case of no error $|\Psi_{00}\rangle$ (and the third error state $|\Psi_{10}\rangle$). This is a powerful tool to read out the desired information about an ensemble in an efficient way. The counter gate is defined as a bilateral controlled-\textit{X} gate, with the qubit pair as the source and the \textit{d}-level system as the target, with action
\be
    \label{eq:acction:CX}
    b\text{CX}^{AB}_{ 1 \to 2} \big| mn \big\rangle_{1} \big| \Psi^{(d)}_{ 0j } \big\rangle_2 = \big|mn\rangle_{1}\big| \Psi^{ (d) }_{ 0, j\ominus m \oplus n } \big\rangle_2.
\ee 
The value $j$ of $|\Psi^{(d)}_{0j}\rangle$ can be obtained by local measurements of $Z_A$ and $Z_B$.

\begin{figure}
    \vspace{-0.3cm}
    \subfloat[]{\includegraphics[width=0.29\columnwidth]{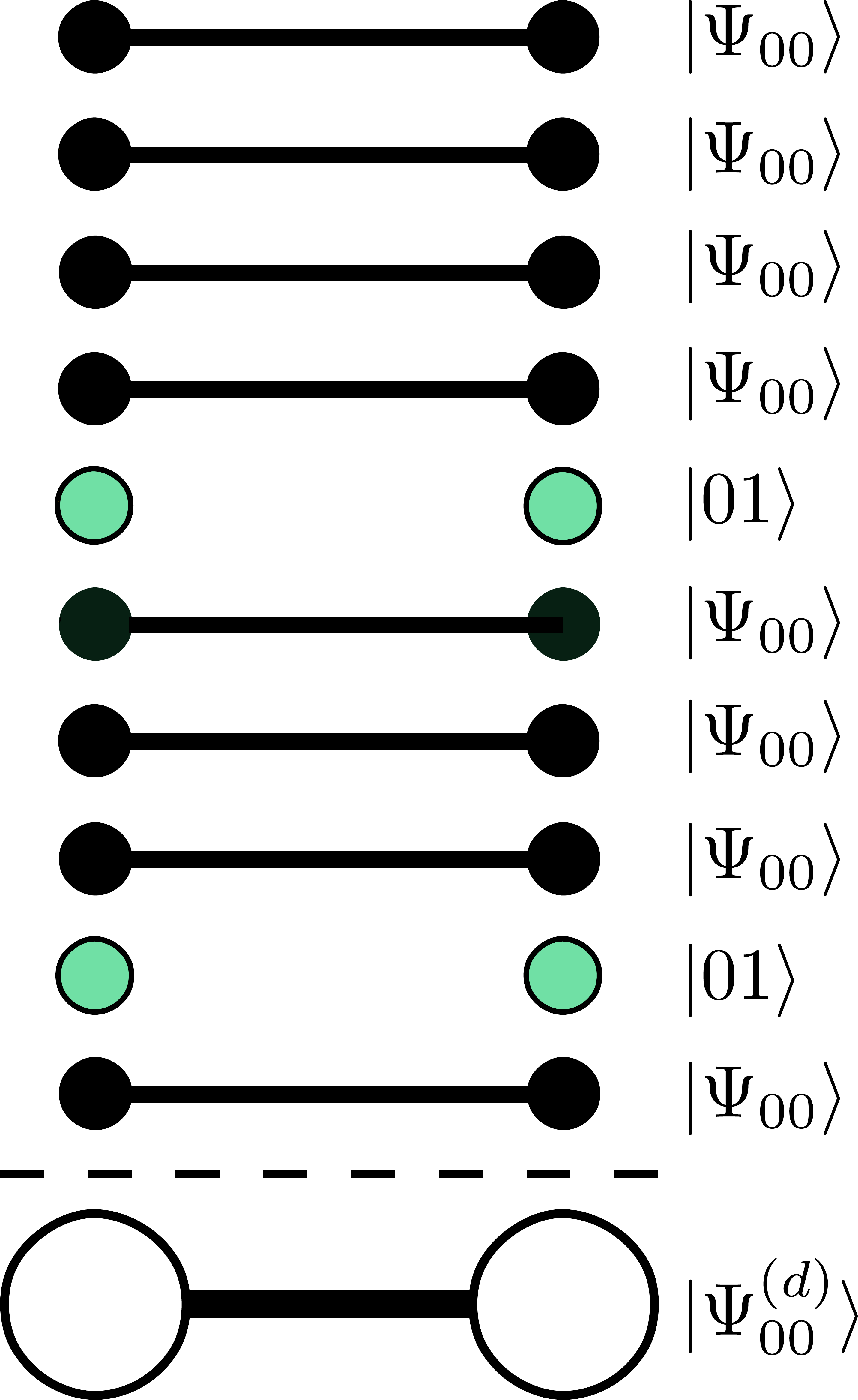}\label{fig:1a}}\hspace{7pt} \subfloat[]{\includegraphics[width=0.67\columnwidth]{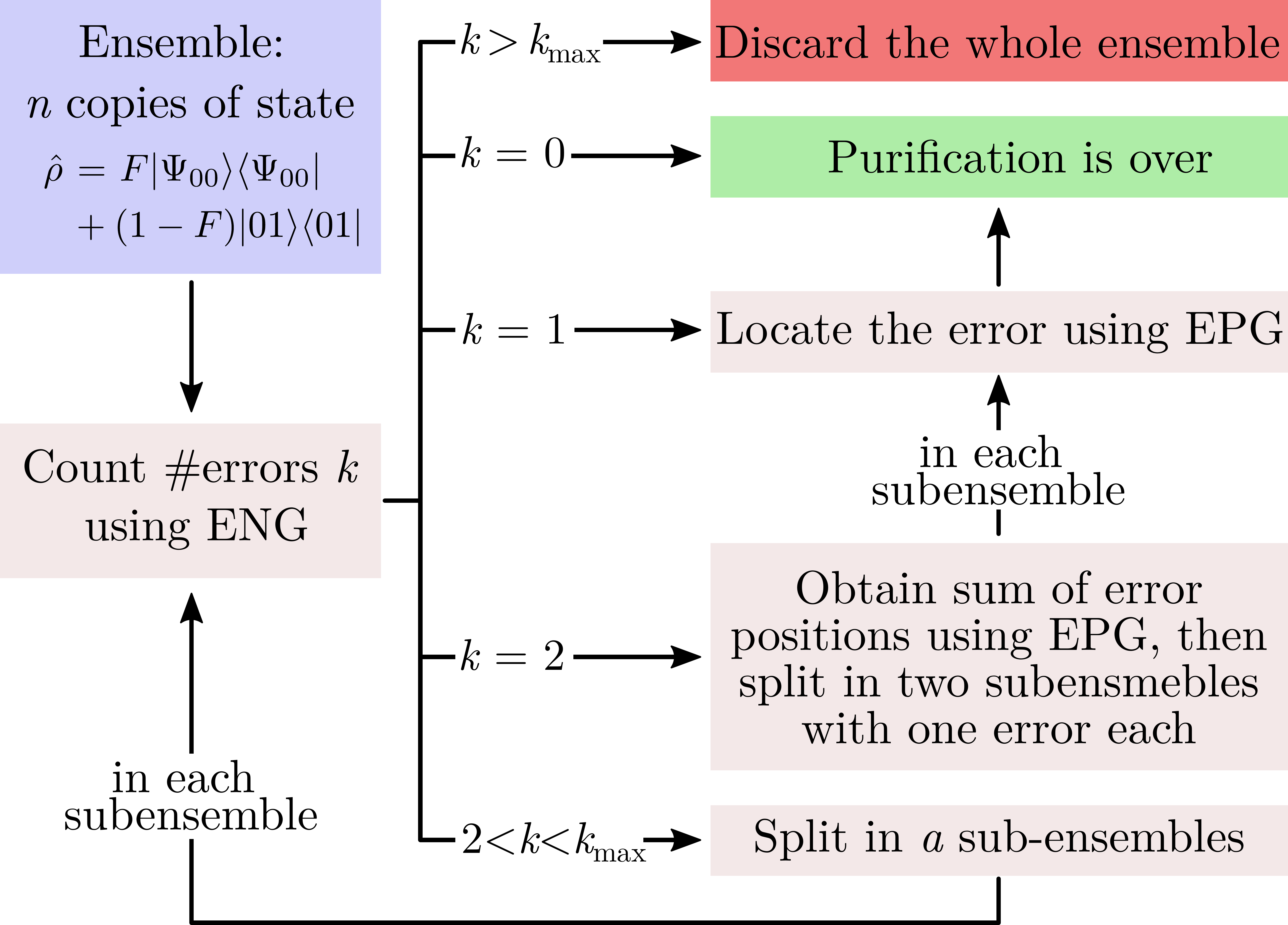}\label{fig:1b}}
    \caption{(a) Schematic representation of an ensemble of rank-2 states in one of the possible configurations. (b) Diagram illustrating the EIP\textsubscript{damp}.}
\end{figure}

{\it Protocol for rank-2 states.---}
The EIP\textsubscript{damp} consists of the following steps: (i) depolarization to standard form Eq.~\eqref{StandardForm}; (ii) determination of the number of errors states in ensemble; (iii) localization of error states. Step (ii) is accomplished by performing the counter gate between each pair in the ensemble and a $d = n+1$-dimensional auxiliary entangled pair, which we call error number gate (ENG). This leaves the auxiliary pair in a state $|\Psi^{(d)}_{0k}\rangle$ if there are $k$ errors. We obtain $k$ with probability $p_k = \binom{n}{k} F^{ ( n - k ) } ( 1 - F )^k$. This result projects the initial ensemble on those configurations with exactly $k$ errors, i.e., all permutations of $\{ |\Psi_{00}\rangle^{ \otimes ( n - k ) } |01\rangle^{\otimes k} \}$. See Fig.~\ref{fig:1a} for a representation of the ensemble. If $k = 0$, we are left with $n$ Bell states, with a conditioned yield of $Y_0  = [ n - \log_2 ( n + 1 ) ] / n$ as the consumed entanglement needs to be returned. Otherwise, we proceed with step (iii). If one error $k = 1$ is found, we perform the counter gate $j$ times for the $j{\rm th}$ pair for all $j$, with the same $d = n$-dimensional auxiliary pair as the target. We call this procedure the error position gate (EPG). In this way, the position of the error state is directly encoded into the target pair. After measuring the auxiliary state we obtain $n-1$ perfect pairs and the conditioned yield of $Y_1 = [ ( n - 1 ) - \log_2 ( n + 1 ) - \log_2 n ] / n$. If $k > 1$, we split the ensemble into two halves and continue with step (ii) until we obtain $k = 0$ or $k=1$ error states in all subensembles, and then proceed with step (iii). See Fig.~\ref{fig:1b} for a diagram summarizing the procedure. This procedure is general and efficient, but can be improved with specific procedures for each $k$. In addition, when the ensemble contains too many errors the required auxiliary entanglement may exceed the purified entanglement, leading to a negative yield for that branch. Therefore, if in step (ii) more than a certain number of errors, $k_{\max}$, are found, one can decide to abort the protocol, discarding the whole ensemble. Therefore, the yield is given by $Y = [ - \log_2 ( n + 1 ) + \sum_{ k = 0 }^{ k_{\max} } p_k \left( n - k - R_k \right) \, ] / n$, where $R_k$ is the number of resources needed to locate all $k$ errors \cite{riera2020purification}.

{\it Approximate EIP for rank-3 states.---}
For states Eq.~\eqref{StandardForm} of rank-3 $(p_3 = 0)$, we have two different error states, with an opposite behavior with the counter gate, see Eq.~\eqref{eq:acction:CX}. Such situation is given, for instance, by bit-flip noise. Consequently, applying the ENG as in step (ii) from above, we do not learn the total number of errors but the difference $k = n_1 - n_2$. Therefore, we are left with an ensemble consisting of permutations of $\{ |\Psi_{00}\rangle^{\otimes n_0} |01\rangle^{\otimes n_1} |10\rangle^{\otimes n_2} \, | \, n_1 - n_2 = k \}$.

As distinguishing between all possible configurations becomes more involved, and requires increasing amounts of auxiliary entanglement, we relax our requirements and allow for errors in the purified ensemble. That is, we consider approximate EIPs that do not aim to output pure Bell states, but only require that the fidelity of the target ensemble exceeds some threshold value. To this aim, we fix some $\lambda\geq 1$ and consider only configurations up to $\lambda$ error states. Configurations with more than $\lambda$ errors will not be detected by our EIP($\lambda$), and hence lower the final fidelity. Notice that the expected number of errors in the ensemble is given by $\lambda_{\rm 0} = (1 - F) n$, and the corresponding binomial probability distribution has almost all of its weight in the interval $\lambda_0 \pm O(\sqrt n)$, which is the basic property of the typical set at the core of standard hashing and breeding protocols \cite{bennett1996mixed,bennett1996purification,dur2007entanglement}. Therefore, for small $n$ we choose a $\lambda > \lambda_0$ sufficiently large to ensure that the probability for configurations with $k > \lambda$ errors is small enough to neglect them. That is what we do in the following. If $n$ is too large, such that $\lambda_0$ is considerable, there is always the possibility of ``blocking'' dividing the ensemble into smaller parts that are purified independently.

{\it EIP$(1)$}: The simplest case is $\lambda = 1$, i.e., we assume there is at most one error in the ensemble. The ENG from step (ii) is applied in the same way as before, but with a $d = 3$-auxiliary register. In this way we distinguish between no error ($k = 0$) and one error of kind $|01\rangle$ ($k = 1$) or $|10\rangle$ ($k = 2 = (-1){\rm mod}\,3$). In case of one error, we apply step (iii) from above to locate it.

{\it EIP$(2)$}: For $\lambda = 2$, we use a $d = 5$-auxiliary state for ENG in step (ii). If we find $k \neq 0$, there is a unique error configuration of exactly one or two errors of a certain kind, and we proceed as in the case of rank-2 states in step (iii). For $k = 0$, there is ambiguity, as it can stem from either no errors ($n_1 = n_2 = 0$) or one error of each kind, $n_1 = n_2 = 1$. To distinguish between the two cases, we use the EPG with an $d = 2n - 1$ auxiliary. This leads to a result $j' = (r - l){\rm mod}( 2n - 1 )$, where $r\,[l]$ is the position of the $|01\rangle$ [$|10\rangle$] error. $j'$ corresponds to the difference of the positions of the two errors, and $j' = 0$ if there are no errors. As detailed in \cite{riera2020purification}, for each $j'$ one can identify two subensembles with one error in each and proceed with (iii) in each subensemble.

In principle, one can devise protocols for larger $\lambda$, and the blocking strategy always allows one to eventually reduce the problem to the case of $\lambda = 1, 2$. In \cite{riera2020purification}, we also describe \textit{a}EIP(3), up to three errors, where for one and two errors the error positions are obtained, while for three errors the protocol aborts since identifying error positions is too costly. The approximate EIP($\lambda$) no longer produce states with unit fidelity, but give states with a fidelity that depends on the particular branch of the protocol one ends up with. The higher the $\lambda$, the larger the reachable fidelity, but the yield is reduced. Notice that one cannot restore perfect entangled pairs, and we use a (virtual) asymptotic hashing protocol to provide conversion rates to estimate yields \cite{riera2020purification}.

{\it EIP for full-rank states.---}
To purify full-rank states, a two-step procedure is required, as the counter gate is not sensitive to $|\Psi_{10}\rangle$ errors. In step I we treat error states $|\phi_1\rangle, |\phi_2\rangle$ as described above. In step II, we then transform the error state $|\phi_3\rangle$ to an equal mixture of $|\phi_1\rangle, |\phi_2\rangle$, while keeping the desired state $|\phi_0\rangle$ unchanged. This is another DEP, given by $\mathcal{D}_2 \circ ( H \otimes H)$, where $H$ is the Hadamard gate. At this point the same procedure as before allows one to detect error states originating from $|\phi_3\rangle$. See \cite{riera2020purification} for details.

\begin{figure}
    \centering
    \subfloat[\centering]{\includegraphics[width=0.9\columnwidth]{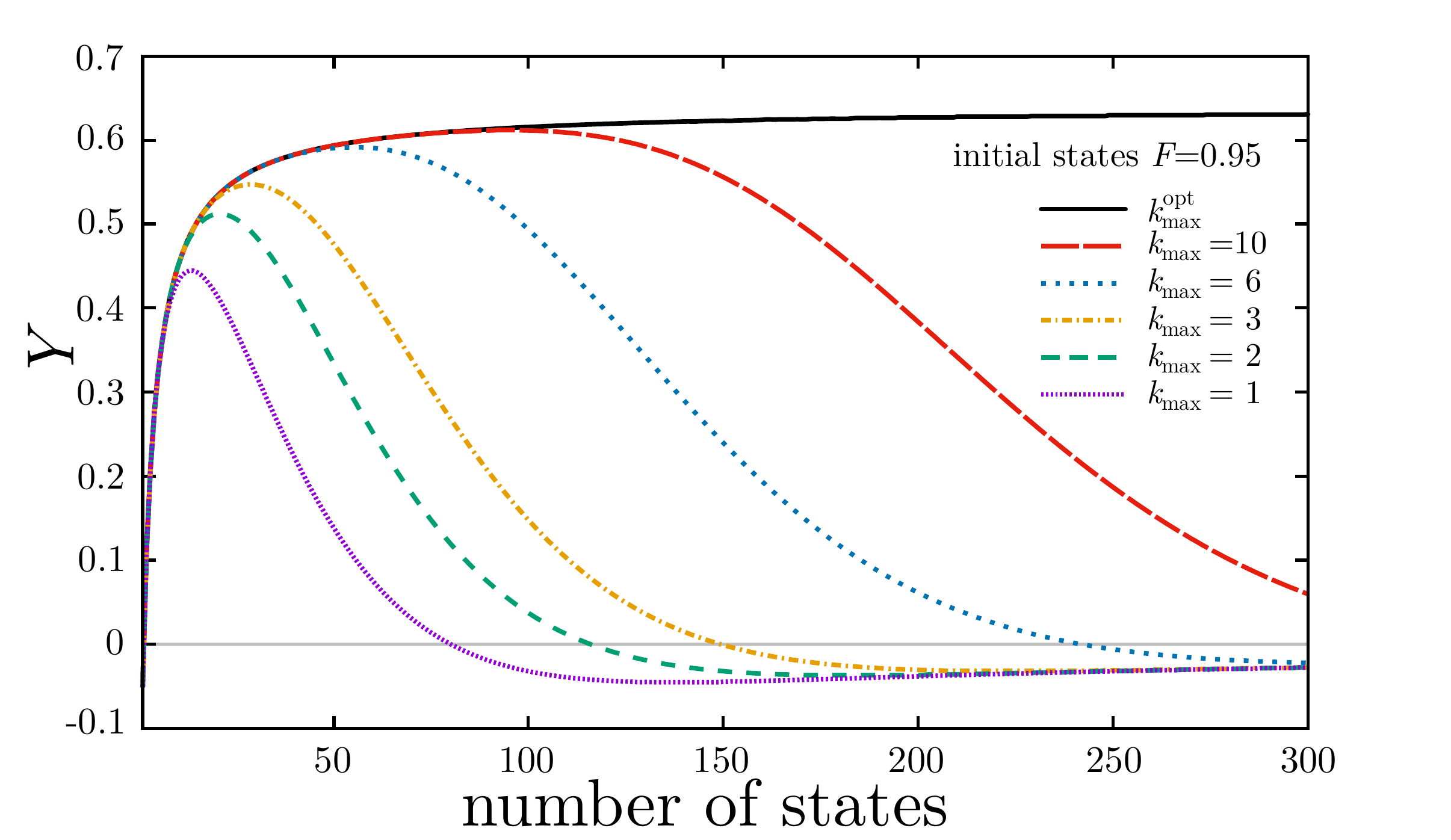}\label{fig:toy}} \hfill
    \subfloat[\centering]{\includegraphics[width=0.9\columnwidth]{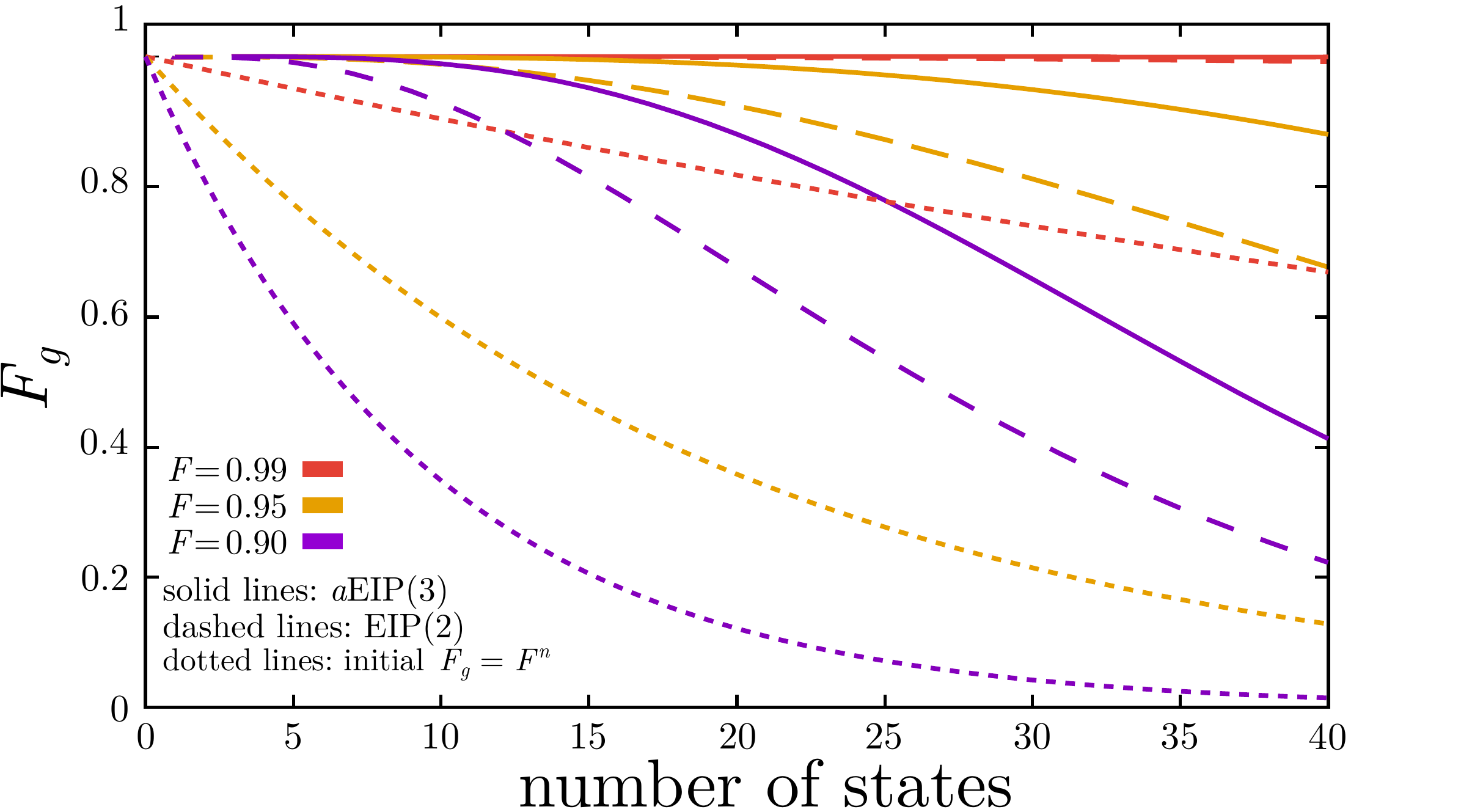}\label{fig:Fg}} \hfill
    \subfloat[\centering]{\includegraphics[width=0.9\columnwidth]{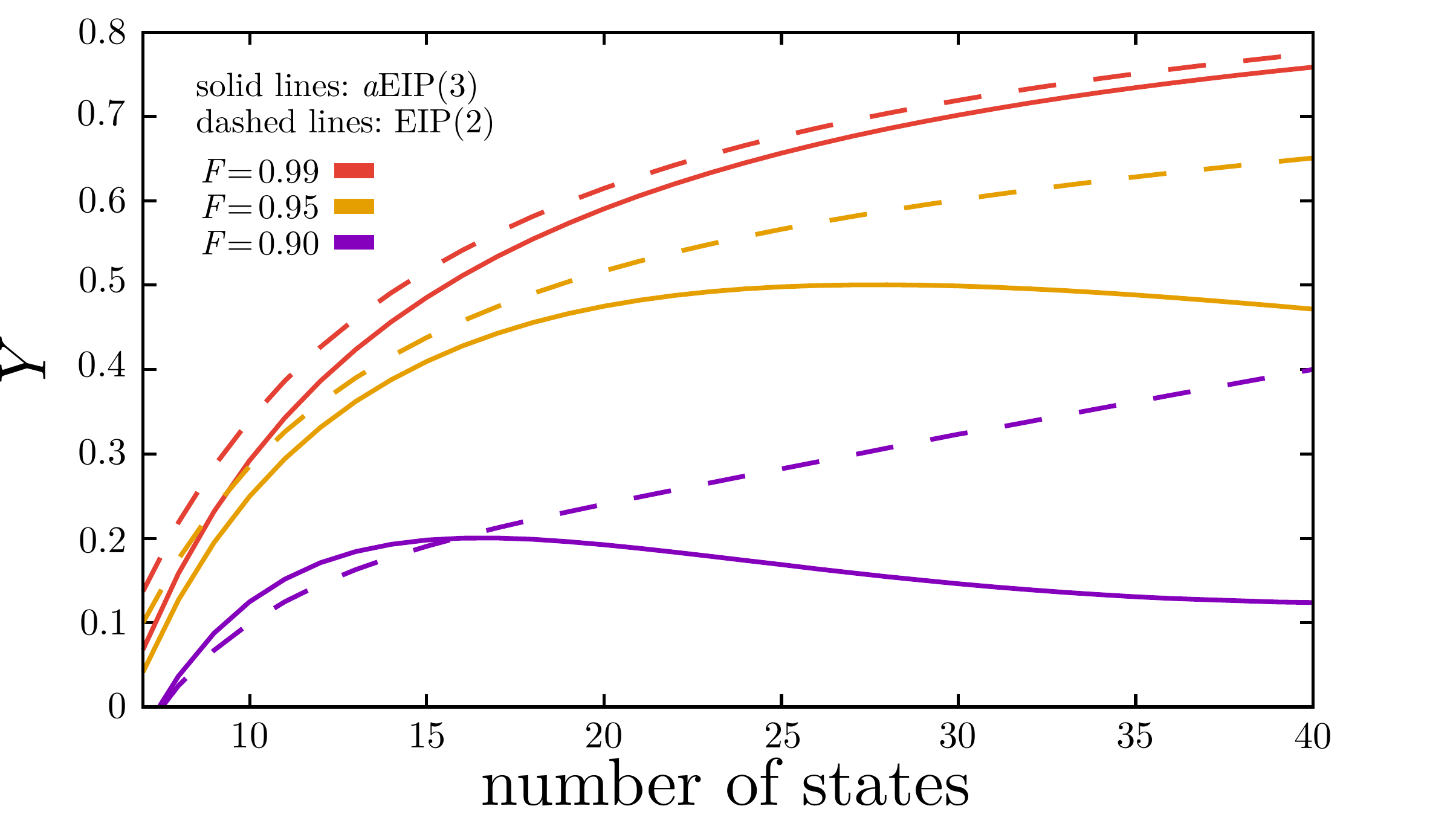}\label{fig:Y}} \hfill
    \subfloat[\centering]{\includegraphics[width=0.9\columnwidth]{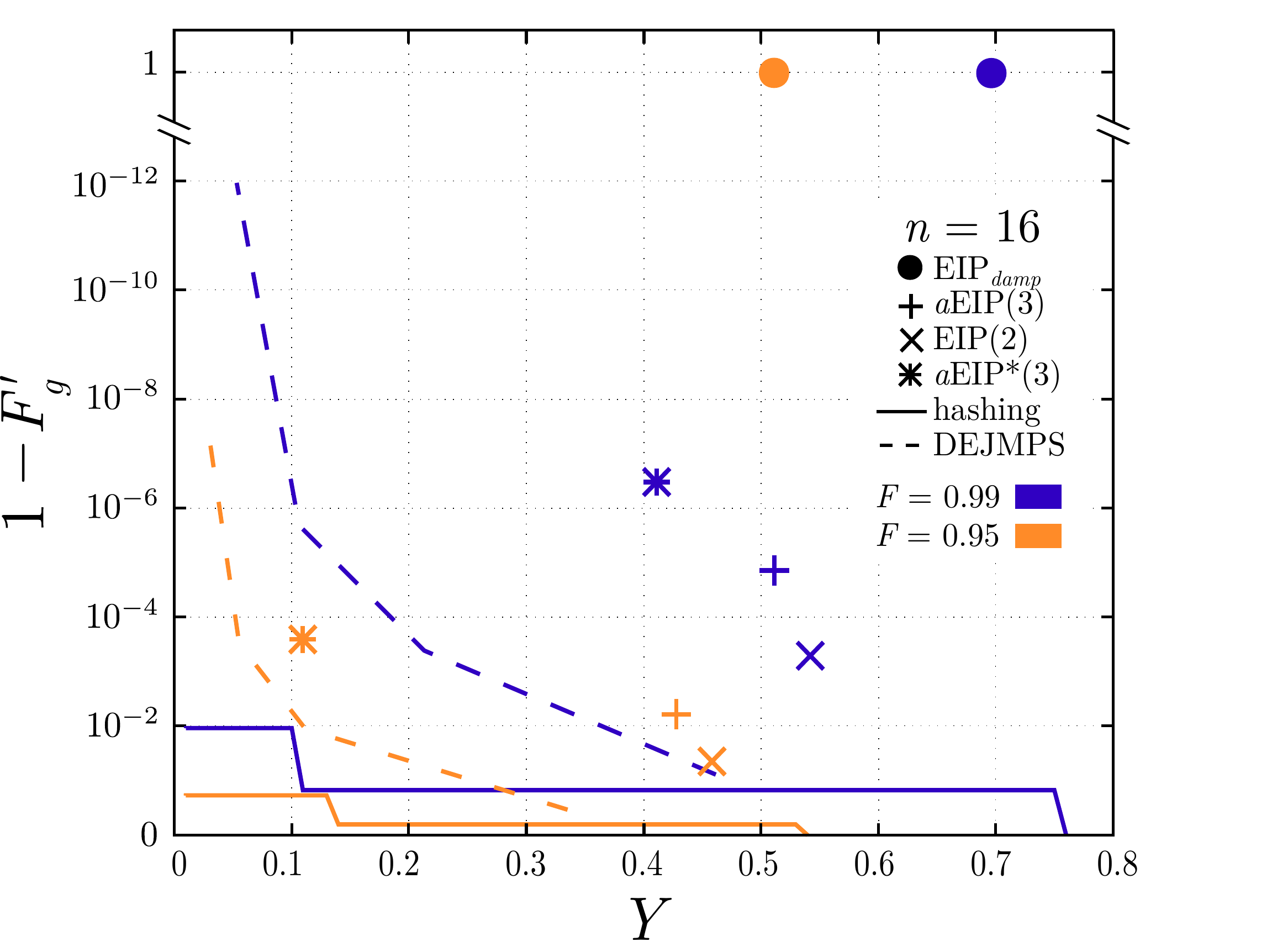}\label{fig:EIPvsDEJHASH}}
    \caption{Yield (a), (c) and global fidelity (b) of different EPP (EIP, dashed lines; \textit{a}EIP, solid lines) for rank-2 input states (a) and rank-3 input states (b), (c) for different initial fidelities as a function of number of initial copies. (d) Comparison of EIP and \textit{a}EIP (asterisk indicates abort whenever error is detected) to recurrence DEJMPS and hashing protocols (upper bound, see \cite{riera2020purification}) in a global fidelity vs yield plot for rank-3 ensembles with $n = 16$ copies (EIP\textsubscript{damp}, circles: rank-2 ensembles, which are equivalent for hashing and DEJMPS protocols).}
\end{figure}

{\it Results.---}
In Fig.~\ref{fig:toy} we show the yield for EIP\textsubscript{damp} (rank-2 states). Fig.~\ref{fig:Fg} and Fig.~\ref{fig:Y} show the reachable global fidelity and yield for EIP(2) and \textit{a}EIP(3) (rank-3 states). In Fig.~\ref{fig:EIPvsDEJHASH}, we compare with finite-size hashing protocols and recurrence protocols \cite{zwerger2018long}. One clearly sees that for moderate system sizes the EIPs perform better than both hashing, breeding, and recurrence protocols.

{\it Discussion.---}
The key element of our EIPs is the readout of locally inaccessible information from a noisy ensemble using auxiliary high-dimensional entanglement \cite{cozzolino2019high,erhard2020advances,wang2020qudits,ecker2019overcoming}. We have put forward concrete schemes to treat particular situations with a fixed number of error states. However, in general, the total number of errors cannot be efficiently obtained which, restricts one to distinguish between situations with at most a certain number of errors. This makes EIPs suitable for ensembles with a small expected number of errors $\lambda_0$. This is appreciated in Fig.~\ref{fig:Fg}, where the output fidelity is close to 1 when the expected number of errors justifies to only distinguish between configurations with at most $\lambda$ errors. Instead, the output fidelity drops when $\lambda_0$ is too large.

On the other hand, the methods we discuss here are not restricted to entanglement purification but have other applications, for instance, to certify the fidelity of a noisy ensemble. Rather than doing state tomography, using entanglement witnesses, or other methods discussed in the literature \cite{Eisert2020}, which all destroy parts of the ensemble, one may use a small amount of auxiliary entanglement to achieve this aim. In such general settings, the entanglement cost of performing nonlocal measurements determines the required resources and, ultimately, the efficiency of the schemes. For ensembles of Bell-diagonal states, we show in \cite{riera2020purification} that deterministically distinguishing between two nonidentical sequences of Bell states requires exactly 1-ebit of entanglement, i.e., an auxiliary Bell state. However, we also find that general dichotomic measurements can require more than 1 ebit \cite{riera2020purification} (e.g., to distinguish the singlet state from the other three Bell states), and the entanglement cost to distinguish configurations that contain Bell states and product states is unknown.

Another relevant aspect is the influence of noise. We have discussed the EIP assuming perfect auxiliary states, and noiseless operations. In principle, one can always use other EPP to purify part of the ensemble and use the resulting high-fidelity states as auxiliary states for the EIP. However, as we demonstrate in \cite{riera2020purification}, EIPs also work using noisy auxiliary states with only $X$ errors. For initial qubit states that are mixtures of only two Bell pairs, e.g., resulting from bit-flip noise, one can generate the high-dimensional auxiliary states from multiple copies of the noisy initial qubit states. For full-rank states one can use the EPP introduced in \cite{deutsch1996quantum} to obtain such kind of noisy Bell states from the initial ensemble. Errors accumulated in the auxiliary state due to local operations can be dealt with in this way, as long as the error rate is sufficiently small as it affects the ensemble. However, EIPs are robust under $X$ errors in the operations, see \cite{riera2020purification} for details.

{\it Conclusions and outlook.---}
We have introduced a class of EPPs based on gathering nonlocal information about an ensemble in a controlled and efficient way. We put forward explicit schemes that work particularly well for ensembles of moderate size where errors result from photon loss or amplitude damping noise. Furthermore, one can extend the protocols for arbitrary noisy initial states and purify large ensembles by blocking them. Alternatively, the possibility of extending our EIPs to locate multiple errors is still open. While we did our analysis for ensembles of the form $\rho^{\otimes n}$, the same protocols are applicable for non-IID situations, as they rely on the determination of the number and position of errors in the ensemble as a whole, regardless of the particular distribution of errors. The protocols can thus also be applied to ensembles where errors are not distributed identically nor independently; only the yield and the final fidelity have to be reevaluated. Given the fact that high-fidelity entanglement is at the heart of most applications in quantum technology, the development of efficient and practical entanglement purification schemes is of crucial importance for their practical realization. For practical reasons, such purification schemes have to be efficient for moderately sized ensembles, at least in the near term. We believe, however, that the techniques and ideas put forward here are not limited to EPP, but have much broader applications. From a practical point of view, establishing methods to perform fidelity estimation or state certification in an entanglement-assisted way is highly relevant, and the counter gate we introduce is a powerful tool in this context. From a more fundamental perspective, the question of entanglement cost of different nonlocal entangling measurements that we put forward and answer partially seems to be worth pursuing further.

\begin{acknowledgments} 
This work was supported by the Austrian Science Fund (FWF) through Project No. P30937-N27 and the Swiss National Science Foundation (SNSF) and the NCCR Quantum Science and Technology, through Grant No. PP00P2-179109 in particular.

We thank Julius Walln\"ofer for interesting discussions.
\end{acknowledgments}

\bibliographystyle{apsrev4-1}
\bibliography{Entanglement-assisted_entanglement_purification.bib}
\end{document}